\documentclass[sigconf]{acmart}
\emergencystretch=\maxdimen
\usepackage{flushend}
\usepackage{balance}
\usepackage{tabu}
\hypersetup{draft}
\AtBeginDocument{%
  \providecommand\BibTeX{{%
    \normalfont B\kern-0.5em{\scshape i\kern-0.25em b}\kern-0.8em\TeX}}}

\usepackage{libertine}

\usepackage[T1]{fontenc}
\usepackage{flushend}

\begin{document}

\pagenumbering{gobble}
\copyrightyear{2020}
\acmYear{2020}
\setcopyright{acmcopyright}
\acmConference[ACMSE 2020]{2020 ACM Southeast Conference}{April 2--4, 2020}{Tampa, FL, USA}
\acmBooktitle{2020 ACM Southeast Conference (ACMSE 2020), April 2--4, 2020, Tampa, FL, USA}
\acmPrice{15.00}
\acmDOI{10.1145/3374135.}
\acmISBN{978-1-4503-7105-6/20/03}
\pagestyle{empty}

\title{Weighted Graph Nodes Clustering via Gumbel Softmax}


\author{Deepak Bhaskar Acharya}
\orcid{}
\affiliation{%
  \institution{Computer Science}
  \streetaddress{5000 Technology Dr NW}
  \city{Huntsville}
  \state{Alabama}
  \country{Usa}
  \postcode{35805}
}
\email{da0023@uah.edu}

\author{Huaming Zhang}
\orcid{}
\affiliation{%
  \institution{Computer Science}
  \streetaddress{5000 Technology Dr NW}
  \city{Huntsville}
  \state{Alabama}
  \country{Usa}
  \postcode{35805}
}
\email{zhangh2@uah.edu}

\begin{abstract}
 Graph is a ubiquitous data structure in data science that is widely applied in social networks, knowledge representation graphs, recommendation systems, etc. When given a graph dataset consisting of one graph or more graphs, where the graphs are weighted in general, the first step is often to find clusters in the graphs.  In this paper, we present some ongoing research results on graph clustering algorithms for clustering weighted graph datasets, which we name as Weighted Graph Node Clustering via Gumbel Softmax (WGCGS for short). We apply WGCGS on the Karate club weighted network dataset. Our experiments demonstrate that WGCGS can efficiently and effectively find clusters in the Karate club weighted network dataset. Our algorithm's effectiveness is demonstrated by (1) comparing the clustering result obtained from our algorithm and the given labels of the dataset; and (2) comparing various metrics between our clustering algorithm and other state-of-the-art graph clustering algorithms.
\end{abstract}

\ccsdesc[500]{Computing methodologies~Cluster analysis}


\keywords{Graph Neural Networks, Gumbel-Softmax, Weighted Graph Clustering.}

\maketitle

\section{Introduction}
Graph clustering concerns the discovery of densely related node groups in a graph. Here an edge between two nodes normally implies any underlying similarity or affinity between nodes, while the absence of an edge suggests dissimilarity and distance. Given the (often noisy) similarity/dissimilarity observations embedded in the network, graph clustering thus attempts to infer groups of closely connected nodes. Graph analysis using the technique of machine learning has been recognized as the strength of graphs is enormous \citep{Keyulu}, i.e., graphs can be used to denote a vast variety of structures in different fields, including \citep{Hamilton} social networks, \citep{Hamaguchi} knowledge graphs, \citep{Khalil} natural science (physical systems \citep{BattagliaPascanu} and protein-protein interaction networks \citep{Shariat}) and many other research areas. In several applications, such as community detection, user profiling, graph clustering, is an essential subroutine. In machine learning, statistics, and pattern recognition, clustering is a subject of active research. Clustering, which is used almost everywhere in computing, is a simple problem. It requires splitting the data into groups of similar objects, and most machine learning applications are the fundamental issue behind it. Data mining leads to datasets with many different attributes being clustered. Therefore the relevant clustering algorithms need particular computational requirements. Several algorithms have recently emerged that meet these requirements and have been used successfully for real-life data mining challenges.

Clustering is the unsupervised mechanism by which natural clusters are found such that objects from the same cluster are identical, and objects from separate clusters are distinct. If similarity relationships between objects are interpreted in clustering as a simple, weighted graph where objects are vertices and similarities between objects are edge weights, clustering reduces the graph clustering problem. However, in many applications, the edge/non-edge between each pair of nodes can reflect very different confidence levels of node similarity. In certain circumstances, edge observation (the absence of an edge) can be created by accurate measurements. Therefore, it is a clear indication that the same clusters should be assigned to the two nodes.In other cases, the findings will be very uncertain, and thus, no knowledge about the cluster configuration will give an edge or the lack of it. The findings between certain node pairs can hold no information as an exceptional example, so they are practically unnoticed.

By extending Acharya and Zhang \cite{acharya2020community} Community Detection Clustering technique via Gumbel Softmax (WGCGS), in this research paper, we propose the new method for Weighted Graph Nodes Clustering via Gumbel Softmax. The article presented by using \cite{acharya2020community} focuses on the unweighted graph datasets. In this research, we recommend finding the cluster of the nodes when the graph is weighted.

The remaining of the paper is organized as follows. In Section \ref{Related work}, we introduce the background and related works in more detail. In Section \ref{Proposed Method}, we introduce our main method. In Section \ref{Experiment Results}, we present the experimental results. In section \ref{Conclusion}, we conclude our work with the future enhancements.

\section{Background and Related work}\label{Related work}
\label{Related work}

\subsection{Community Detection Clustering via Gumbel Softmax}
Different algorithms have been proposed, and there is extensive research on community detection. Detecting communities is a method for defining related groups that can be difficult, depending on the graph network's size and scale. Acharya and Zhang \cite{acharya2020community} provide a general understanding of their paper on working with unsupervised graph datasets and identifying graph nodes' clusters.

In general, let the adjacency matrix be $A_{n\times n}$, where 'n' represents the total number of nodes in the graphs dataset. The $W_C$ matrix size is $n\times k$, where k indicates the number of clusters. After obtaining a matrix of the size $k \times k$ then perform the $W_C^tAW_C$ operation and call this matrix as $R_{k \times k}$. 

The resulting $R_{k \times k}$ matrix displays the cluster strength, where the primary diagonal shows the strength of the data within the cluster groups, and the other $R$ matrix components show the strength of the data between the different clusters. Then as a discrete distribution of probabilities, apply the softmax function to the $R$ matrix obtained to express inputs. Mathematically be defined as follows:

\begin{equation} \label{actual_gumbel}
Softmax(x_i)=\frac{exp(x_i)}{\sum_{j=0}^{m}{exp(x_j)}} \ for \ i = 1,2,3, ...... ,m.
\end{equation}

\subsection{Gumbel Softmax Approach on Graph Features Selection and Extraction}
In the suggested approach for the citation datasets, the article features selection and extraction for Graph Neural Networks, Acharya and Zhang \cite{bhaskar2019feature} selected and extracted Graph Neural Network (GNN). By applying the feature selection and extraction technique to GNNs using gumbel softmax, they conduct tests using different reference datasets: Cora, Pubmed, and Citeseer. Acharya and Zhang demo an example of the Citeseer dataset, where 375 of 3703 features are selected and ranked according to prominent features. The proposed deep learning method works well with reduced features, decreasing the number of features that the dataset initially had by about 80-85\ percent. The experiment results show that the accuracy declines steadily when using the selected classification features falling within the ranges 1-75, 76-150, 151-225, 226-300, and 301-375.

In general, let the input feature matrix be $X_{n \times f}$ where 'n' and 'f' represent the total number of nodes and features in the graph's dataset. The resulting matrix is $M_{n \times k}$, where 'k' represents the features we select from the 'f' features.

Acharya and Zhang select the features in a graph citation dataset using the gumbel softmax function and the proposed approach. Gumbel softmax's distribution is \cite{Shixiang, maddison2016concrete}' a continuous distribution over the simplex that can approximate categorical distribution samples.' A categorical distribution is a one-hot vector that defines the maximum probability of one and all the other probability as zero.

The two layer Graph Convolution Network (GCN) used in the experiment is defined as 
\begin{equation} \label{gcn_1}
GCN(X,A) = Softmax(A (ReLu(AXW_GW_1)) W_2) 
\end{equation}
To verify the selected features and calculate the accuracy for classification they use the following two layer Graph Convolution Network as defined below
\begin{equation} \label{gcn_2}
GCN(X,A) = Softmax(A (ReLu(AXW_G')) W_2) 
\end{equation}
$A$: Adjacency matrix of the undirected graph G.  \newline
$X$: Input feature matrix. \newline
$W_G$: gumbel softmax feature selection / feature extraction matrix. \newline
$W_G'$: feature selection / feature extraction matrix obtain
ed from the result of Equation \ref{gcn_1}.\newline
$W1,W2$: Layer-specific trainable weight matrix.\newline
$ReLu$ : Activation function ReLu(.) = max(0,.).

\section{Proposed Method}\label{Proposed Method}
The proposed approach in the article Community Detection Clustering via Gumbel Softmax (CDCGS) clusters of graph dataset nodes and the article Feature Selection and Extraction for Graph Neural Networks (FSEGNN) explains how prominent features for graph citation datasets can be selected and extracted. The dataset used is a graph in each of the papers. There are nodes and edges in the graphs, and they have relations between the nodes that can have some knowledge about the similarities or dependencies between the nodes. The two papers above do not provide insight into the network of weighted graphs. We suggest a new method of Weighted Graph Nodes Clustering via Gumbel Softmax (WGCCS) to cluster the graph nodes as an extension to the CDCGS and FSEGNN.

We consider the network of 'n' nodes and the adjacency matrix 'A' in our WGCCS method. An adjacency matrix is used to represent a finite graph. The elements of the matrix show whether the pairs of vertices in the graph are adjacent or not. Then we cluster the graph into k clusters, applying our method.
 
Gumbel softmax distributions can be used to estimate the sampling process of discrete data if we have a stochastic neural network with discrete variables. Using backpropagation, the network can then be trained, where the efficiency of the network will depend on the temperature parameter choice.

The method used in the experiment is defined as below:
\begin{equation} \label{gcn_2}
Weighted-Graph-Cluster(Adj) = Softmax(W_C^t(Adj)W_C) 
\end{equation}

In the Equation \ref{gcn_2}, '$Adj$' indicates the Adjacency matrix of the undirected weighted graph G, where weight of graph edges indicates the similarity between the nodes of the graph and '$W_C$' indicates the gumbel cluster weight matrix.

We have the gumbel cluster weight matrix $W_C$ of size $N\times k$ once model training is completed, where $N$ is the number of graph nodes, and $K$ is the number of clusters. Then, we look at the sum of the row elements, where each row sums up to 1, and the row number's maximum index is the cluster to which the graph node belongs. For example, let's consider k= 2, i.e., we're trying to cluster the dataset into two cluster groups. Let us say that the data is [0.65 0.35] for the first row, where 0.65 is $0^{th}$, and 0.35 is $1^{st}$. By looking at the maximum value in the row, we get the corresponding maximum value index, and one can easily say that the data belongs to cluster 0. If [0.39 0.61] is the data for the second row, 0.39 will be $0^{th}$, and 0.61 will be $1^{st}$ index. By looking at the maximum value in the row, we get the corresponding maximum value index, and one can easily say that the data belongs to cluster 1.

The resulting ${k \times k}$ size matrix derived from the \ref{gcn_2} equation is compared to the identity matrix ($I_{k \times k}$) to find network losses. Here the cluster groups represent diagonal components. We may then form the cluster groups of the data points by using our proposed approach. A detailed analysis is given in the result section, taking into account an example of a dataset weighted karate club network.

\section{Experiment Results}\label{Experiment Results}
    \subsection{Datasets}
    Zachary's Weighted Karate Club Network is a well-known dataset describing the relationships in a university karate club used by Wayne W. Zachary in his paper "An Information Flow Model for Conflict and Fission in Small Groups." This dataset is known for its simple description of the community structure, which occurs because it is possible to cluster network nodes into strongly linked sets. Focused on Mr. Hi, the karate instructor, and John A, the club president, Zachary's Karate Club network can be split into two groups. The network accurately predicts how, following a disagreement over the pay, the karate club splits into two new clubs and creates a rift within Hr. Hi and John A. The network shows which members of the club will join the new club by analyzing community members' meetings outside the club's background in 33 of 34 instances.
   
\begin{figure}[!htbp]
    \centering
    \begin{minipage}{\textwidth}
      \includegraphics[scale=0.35]{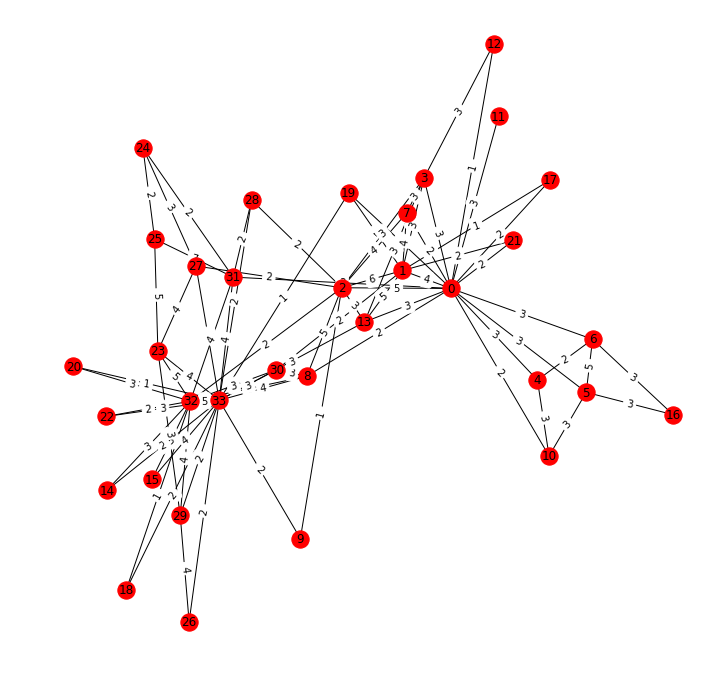}
    \end{minipage}\hfill
    \caption{Weighted Karate Club Network} 
    \label{karate}
\end{figure}

   The initial dataset of Zachary's Karate Club is weighted by multiple friendship measures. Several visualizations of Zachary's Karate Club have been created since the first article was published in 1977. Michelle Girvan and Mark Newman \cite{girvan2002community} used Zachary's Karate Club again in 2002 to demonstrate community structure in their article "Community structure in social and biological networks." Figure \ref{karate} is a network created by a computer that contains each member of the club.
   
    \subsection{Results}
\begin{figure}[!htbp]
    \centering
    \begin{minipage}{\textwidth}
      \includegraphics[scale=0.35]{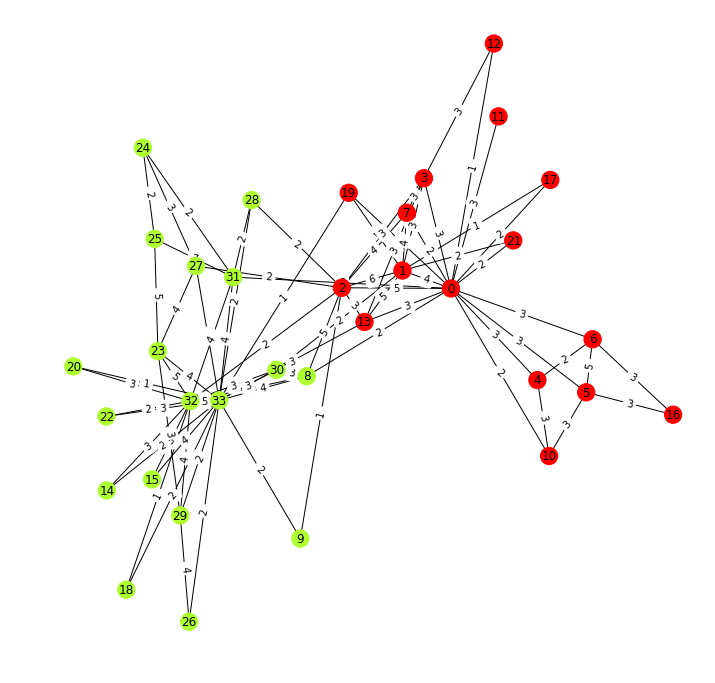}
    \end{minipage}\hfill
    \caption{Weighted Karate Club Network with 2-Clusters} 
    \label{karate2}
    \end{figure}

    The weighted network of Zachary karate club is the network we can think of to evaluate our methodology. We also know that the two teacher nodes are dominant graph nodes 0 and 33, as the weighted Zachary network is labeled.  Any other network nodes can either join the teacher community of node 0 or the teacher community of node 33. Until clustering, the original network is seen in figure \ref{karate2}. Second, we run our strategy to cluster the network into 2-clusters like the Zachary karate club's ground-truth community structure has 2-clusters.  We compare our proposed method results with the results of algorithms such as greedy optimization of leading eigenvector of the community matrix(ECM) \cite{newman2006finding}, edge betweenness(EB) \cite{PhysRev}, short random walks(RW) \cite{pons2005computing}, Infomap community finding(ICF) \cite{Rosvall_2009}, multi-level optimization of modularity(MOM) \cite{Blondel_2008}, propagating labels(PL)  \cite{Raghavan_2007}. Metrics such as adjusted randomized index (ARI) \cite{sklearn_api}, homogeneity (HOMO), normalized mutual information (NMI), completeness (COMP), and v-measure V-MES) \cite{rosenberg2007v} has also been computed to complete the analysis. ARI = 1, HOMO = 1, NMI = 1, COMP = 1, V-MES = 1 and the modularity calculation = 0.371 are the corresponding metric metrics if we cluster the weighted Zachary network into 2-clusters (WGCGS(2C)) using WGCGS. The value of ARI, HOMO, NMI, COMP, and V-MES indicates that the proposed algorithm clusters the nodes to their respective groups as per the target label. Figure \ref{karate2} represents Zachary's network of 2-cluster karate clubs resulting from our approach proposed.
    
\begin{figure}[!htbp]
        \centering
        \begin{minipage}{\textwidth}
          \includegraphics[scale=0.35]{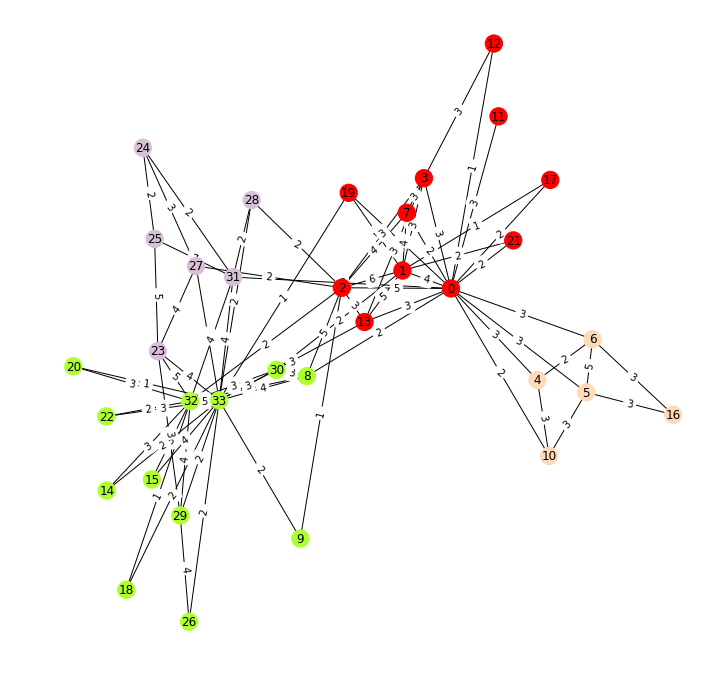}
        \end{minipage}\hfill
        \caption{Weighted Karate Club Network with 4-Clusters} 
        \label{karate4}
    \end{figure}

     Whenever we consider the graph dataset, the higher the modularity value indicates a stronger assignment to the cluster. To obtain higher modularity for the stronger assignment, we clustered the weighted Zachary karate club network into four clusters. 4-clustered Zachary network (WGCGS(4C)) gives us a 0.4197 modularity metric (Figure \ref{karate4}). Considering that students joining node 0 as cluster group 1 and node 33 as cluster group 2, node 0 and node 33 are already connected to several nodes, and nodes 23, 24, 25, 27, 28, 31 (cluster 3) and 4, 5, 6, 10, 16 (cluster 4) mean that students are not joining cluster group 1 or 2. To test the classification of metadata groups by various algorithms, we used the best scores of the Zachary karate club network along with different algorithms. Our WGCGS approach outperforms the metric results of the ECM, EB, RW, MOM, and PL algorithms, and we presented them in Table \ref{Table_result2}. 
    
   The result obtained from the \ref{gcn_2} equation is compared to the loss function. The loss function used for our experiment is the identity matrix ($I_{k \times k}$), where the cluster group is represented by each diagonal member.
    
\begin{table*}
     \caption{Zachary karate club network metrics for the different algorithms and our method (WGCGS) }
        \label{Table_result2}
        \begin{tabu} to \textwidth {XXXXXXXXX}
            \toprule
           \textsc{\textbf{Metric}} & 
            \textsc{\textbf{ECM}}&
            \textsc{\textbf{EB}}&
            \textsc{\textbf{RW}} & 
            \textsc{\textbf{ICF}} & 
            \textsc{\textbf{MOM}} &
            \textsc{\textbf{PL}} &
            \textsc{\textbf{WGCGS(2C)}}&
            \textsc{\textbf{WGCGS(4C)}}\\
            \midrule
            ARI    & 0.5120 & 0.5125      & 0.2620 &0.7021 &0.4619 &0.4714 &0.5414 &1\\
             NMI    & 0.6770 & 0.6097      & 0.3729 &0.6994 &0.5866 &0.5282 &0.6873 &1\\ 
              HOMO   & 1 & 0.8850       & 0.5588 &0.8535 &0.8535 &0.6905 &1 &1\\ 
              COMP   & 0.5118 & 0.4651     & 0.2798 &0.5925 &0.44.68 &0.4277 &0.5235 &1\\ 
              V-MES   & 0.6770 & 0.6097     & 0.3729 &0.6994 &0.5866 &0.5282 &0.6872 &1\\ 
            \bottomrule
        \end{tabu}
    \end{table*}

\section{Conclusion and Enhancements}
\label{Conclusion}
We introduced the Weighted Graph Nodes Clustering via Gumbel Softmax strategy for the weighted karate club network. The experimental results illustrate the efficiency by selecting appropriate parameter values and checking the resulting clustering accuracy. The approach currently available is just as diverse as the implementations for weighted undirected graph clustering. We are currently experimenting with applying the principle of clustering for Graphs on few more weighted graph datasets and a directed graph dataset.

\bibliographystyle{abbrv}
\bibliography{bib-sam}


\end{document}